\documentclass[twocolumn]{aastex701}
\usepackage[utf8]{inputenc}
\usepackage{graphicx}
\usepackage{amsmath}
\usepackage[hyphens]{url}
\usepackage{hyperref}

\newcommand{\ki}{\mathbf{k^i}}
\newcommand{\robs}{R_\mathrm{obs}}

\submitjournal{The Astrophysical Journal}
\received{Deceber 8, 2025}
\revised{February 22, 2026}
\accepted{February 28, 2026}

\shorttitle{Hot Jupiters Inflated Mainly by Shallow Heating}
\shortauthors{Schmidt, Thorngren, and Schlaufman}

\begin{document}
\title{Hot Jupiters are Inflated Primarily by Shallow Heating}
\author[0000-0001-8510-7365]{Stephen P.\ Schmidt}
\altaffiliation{NSF Graduate Research Fellow}
\affiliation{William H.\ Miller III Department of Physics \& Astronomy,
Johns Hopkins University, 3400 N Charles Street, Baltimore, MD 21218, USA}
\email{sschmi42@jhu.edu}

\author[0000-0002-5113-8558]{Daniel P.\ Thorngren}
\affiliation{William H.\ Miller III Department of Physics \& Astronomy,
Johns Hopkins University, 3400 N Charles Street, Baltimore, MD 21218, USA}
\email{dpthorngren@jhu.edu}

\author[0000-0001-5761-6779]{Kevin C.\ Schlaufman}
\affiliation{William H.\ Miller III Department of Physics \& Astronomy,
Johns Hopkins University, 3400 N Charles Street, Baltimore, MD 21218, USA}
\email{kschlaufman@jhu.edu}

\correspondingauthor{Stephen Schmidt}
\email{sschmi42@jh.edu}

\begin{abstract}
\noindent
The unexpectedly large radii of transiting hot Jupiters have led to many proposals for the physical mechanisms responsible for heating their interiors.
While it has been shown that hot Jupiters reinflate as their host stars brighten due to heating deep in planetary interiors, young hot Jupiters also exhibit signs of delayed cooling possibly related to heating closer to their surfaces.
To investigate this ambiguity, we enhance our previously published hot Jupiter thermal evolution model by adding a parameter that allows for both deep heating and delayed cooling.
We fit our thermal evolution models to a homogeneous, physically self-consistent catalog of accurate and precise hot Jupiter system properties in a hierarchical Bayesian framework.
We find that hot Jupiters' interior cooling rates are reduced on average by 95\%--98\% compared to simpler anomalous heating models.
The most plausible explanation for this inference is substantial shallow heating just below their radiative--convective boundaries that enables reinflation with much weaker deep heating.
Shallow heating by Ohmic dissipation and/or temperature advection are therefore important components of accurate models of hot Jupiter atmospheres, especially in circulation models.
If hot Jupiters are inflated primarily by shallow heating as we propose, then we predict that atmospheric circulation-related observables should increase with temperature in the range $T_{\text{eq}}~\lesssim1500~\text{K}$, peak in the range $1500~\text{K}~\lesssim~T_{\text{eq}}~\lesssim~1800~\text{K}$, and decrease in the range $T_{\text{eq}}~\gtrsim~1800~\text{K}$.
\end{abstract}


\section{Introduction} \label{sec:Introduction}
The radii of giant exoplanets that orbit very close to their host stars appear consistently larger than interior structure models would predict.
These ``hot Jupiters'' have sufficiently small orbital separations that their strong instellation lengthens the atmospheric scale height and slows the loss of heat.  It was expected that these planets would be somewhat inflated relative to the gas giants in our solar system.
Indeed, even the first known transiting hot Jupiter, HD 209458 b, was found to be larger than Jupiter despite being $\sim70$\% its mass \citep{Henry2000, Charbonneau2000}. 
However, thermal evolution models of these planets struggled to match the substantially larger radii observed without introducing a powerful interior heat source \citep[e.g.,][]{Bodenheimer2001, Guillot2002, Baraffe2003}.
As more hot Jupiters have been discovered via ground-based transit surveys like the Wide Angle Search for Planets \citep[WASP;][]{Pollacco2006, CollierCameron2007}, Hungarian Automated Telescope Network \citep[HATNet;][]{Bakos2004, Bakos2007}, Kilodegree Extremely Little Telescope \citep[KELT;][]{Pepper07}, and Next Generation Transit Survey \citep[NGTS;][]{Wheatley18}, as well as space-based transit surveys like the ESA's CoRoT \citep{Baglin06} mission and NASA's Kepler \citep{Borucki2010}, K2 \citep{Howell2014}, and Transiting Exoplanet Survey Satellite \citep[TESS;][]{Ricker2014} missions, individual cases of larger-than-expected radii have developed into a clear trend for planets with $T_\mathrm{eq} > 1000$\;K \citep[e.g.,][]{Miller2011, Demory2011, Thorngren2016}.
This discrepancy is now referred to as the hot Jupiter radius inflation problem\footnote{See \citet{Thorngren2024} for a recent review of the topic.}, which asks why hot Jupiters remain large and internally hot for so long after they form.

Statistical analyses of the hot Jupiter population have revealed some details about the nature of this problem. It was expected almost immediately after their initial discovery that hot Jupiter radius inflation should depend on some combination of a planet's irradiation, mass, and age \citep[e.g.,][]{Guillot2002, Burrows2004}. As the population of hot Jupiters reached a size suitable for statistical analyses, hot Jupiter orbit-averaged equilibrium temperatures were found to correlate with the discrepancy between their measured radii and \citet{Bodenheimer03} non-anomalously heated theoretical models \citep{Laughlin11}. The stellar flux experienced by a hot Jupiter was subsequently identified as a far better predictor of a planet's radius than its orbital period \citep{Weiss2013}. The effort to generate an empirical relation between planet radius and other independent variables was expanded upon by \citet{Thorngren2021}, who found that incident flux and planet mass are strongly favored as predictors of planet radius, with a marginal additional dependence on stellar atmospheric metallicity.

The observed correlation between incident flux and planet radius identified by \citet{Weiss2013} suggests that planets may grow in size as their parent stars brighten \citep{Lopez2016}.  \citet{Grunblatt2016, Grunblatt2017} found evidence for such reinflation around post main-sequence stars.  Additionally, \citet{Hartman2016} inferred similar reinflation with stellar evolution on the main sequence.  \citet{Thorngren2021} reproduce this result and show it is a plausible outcome of planetary thermal evolution; however, this finding was complicated by a negative correlation between planet radius and age though. This was explained as a consequence of either more massive stars having shorter main sequence lifetimes than less massive stars or an additional mechanism that delays hot Jupiter cooling.
The presence of more than one cause of unexpectedly large radii complicates the picture, requiring more complex models to explain observed radius inflation.

A primary differentiator between hot Jupiter heating mechanisms is their depth in the envelope, as this determines its effect on a planet's radius evolution over time \citep{Komacek2020}.
Some mechanisms, like Ohmic dissipation \citep[Lorentz forces on particles driven by winds through a magnetic field; ][]{Batygin2010, Batygin2011, Perna2010a}, compositional gradients \citep[e.g.,][]{Chabrier2007}, or fluid dynamical effects like turbulent mixing \citep{Youdin2010} and radial potential temperature advection driven by superrotating longitudinal jets and other flows \citep{Tremblin2017}, are likely only able to deposit heat in the shallow interior.
Such mechanisms make the planet's temperature-pressure profile shallower, limiting the efficiency of outward heat escape and slowing radius contraction.
However, as convection must cause a net upward heat flux, the only avenue for reinflation is the much slower conduction.
Shallow heating therefore takes tens of Gyr to reinflate the planet --- far too long to be observable.

Other mechanisms like tidal dissipation \citep[e.g.,][]{Bodenheimer2001, Leconte2010} or thermal tides \citep[i.e., atmospheric temperature gradients causing tidal distortion in an attempt to reach equilibrium as suggested by][]{Arras2009, Arras10, Socrates2013} may primarily occur deep in the interior.
These mechanisms heat the planet throughout its interior, but do not slow the rate of heat escape from the interior.  As such they can reinflate planets and keep them large, but would not produce the observed delayed cooling trends \citep[e.g.,][]{Thorngren2021}.
To resolve this tension, we will model the observed hot Jupiter population as having both deep heating and delayed cooling\footnote{To clarify, here we define hot Jupiter ``heating models'' as models that incorporate ``mechanisms'' that cause the planet's radius to be ``inflated,'' i.e., larger than what would be expected without anomalous heating. While typically ``inflation'' would imply that the planet became larger over time, we use it as a catchall term for both ``reinflation,'' which follows the typical definition, and ``delayed cooling,'' which occurs in the opposite direction but still causes unexpectedly large radii.}.

We will use a population-level analysis of the hot Jupiter population's radius inflation to understand the relative contributions of deep heating and delayed cooling.
By applying giant planet thermal evolution models to a hierarchical Bayesian inference framework, we will be able to use individual planet parameter inferences to infer a population-level value.
This methodology has been used previously in similar circumstances: \citet{Thorngren2018} applied it to the population of hot Jupiters known at the time to extract the relation between heating efficiency and incident flux.
Now we employ it with more advanced thermal evolution models.

This analysis requires a homogeneous, physically self-consistent sample of planet and host star properties as an inhomogeneous sample of stellar masses, radii, and ages could bias the inferred cooling rates.
In particular, precise ages are valuable for constraining the rate at which hot Jupiters cool.
Age inference for mature exoplanet host stars remains difficult in the absence of asteroseismic or stellar mean density constraints from high-precision light curve data.
However, our hierarchical Bayesian inference methodology mitigates the effects of limited information from a single system by aggregating results from many systems.
This allows us to obtain a precise inference as attained by other population-level exoplanet age studies \citep[e.g.,][]{Hamer2019, Schmidt24}.

In this article we will extract the population-level contributions of deep heating and delayed cooling to hot Jupiter radius inflation.
We introduce our thermal evolution model in Section \ref{sec:thermalEvolutionModel}.
We describe the construction of our homogeneous sample of hot Jupiter system properties in Section \ref{sec:stellarInference} and our application of these data to a hierarchical Bayesian inference framework in Section \ref{sec:hierarchicalBayesianInference}.
We show our results and discuss their implications in Section \ref{sec:Discussion}, and summarize them Section \ref{sec:Conclusions}.

\section{Thermal Evolution Model}\label{sec:thermalEvolutionModel}
\begin{figure*}[t!]
    \centering
    \includegraphics[width=\linewidth]{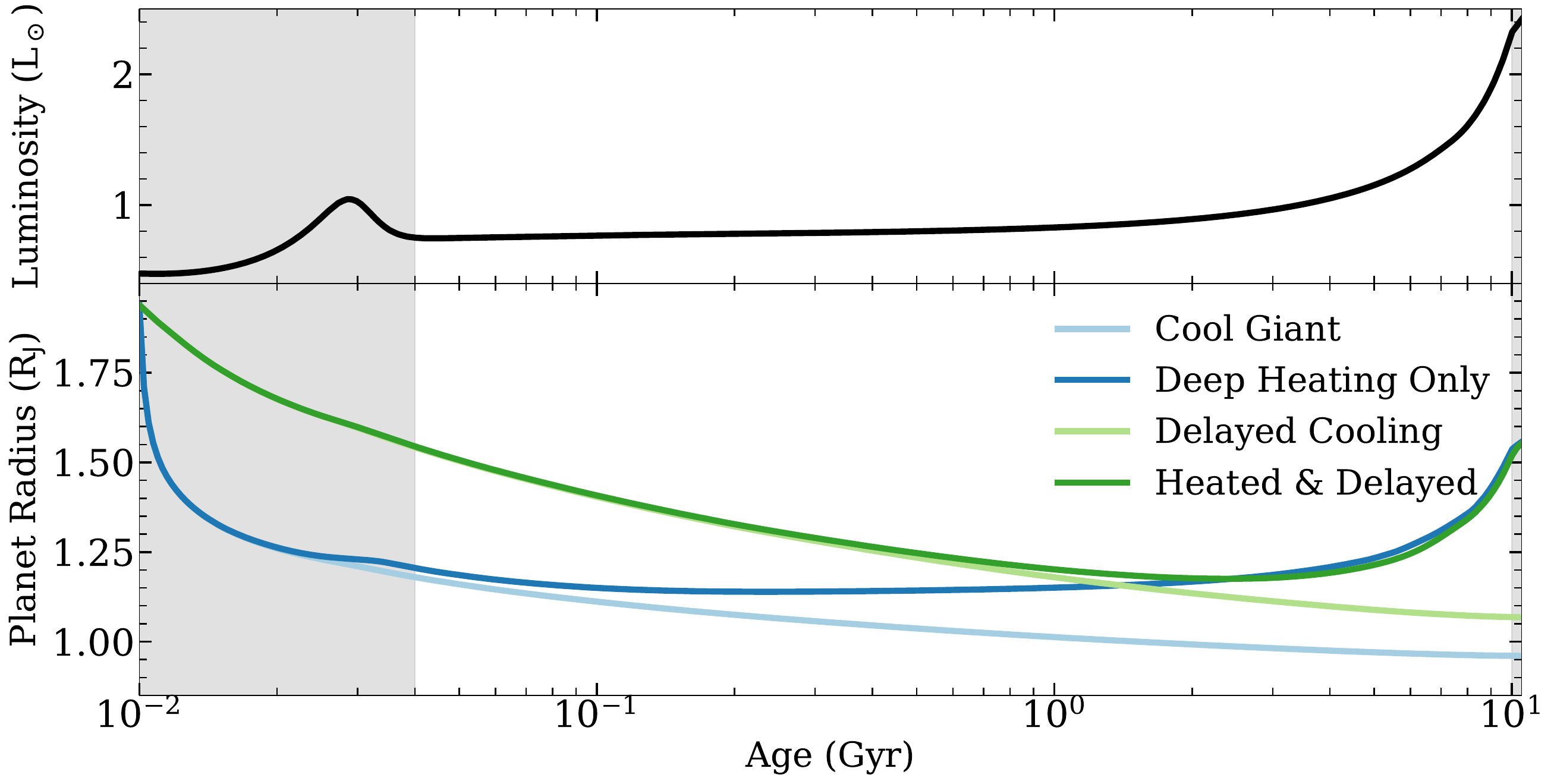}
    \caption{Demonstration of our thermal evolution model's interior heating parameterization and its effects on the modeled planet's radius evolution. In the top panel, we plot as the black line a 1 M$_\odot$ host star's luminosity as a function of age, highlighting the pre- and post-main sequence region as the shaded areas. In the bottom panel, we plot as colored lines the 1 M$_{\text{Jup}}$ planet's radius assuming (1) standard thermal evolution (light blue), (2) only deep interior heating (dark blue), (3) only delayed cooling (light green), and (4) both deep heating and delayed cooling (dark green). Deep heating allows the planet to reinflate as the host star's luminosity increases, while delayed cooling slows the rate of contraction.}
    \label{fig:thermalModel_heating}
\end{figure*}

Our planetary interiors are modeled similarly to \citet{Thorngren2023}; we solve in one dimension the equations of hydrostatic equilibrium, mass conservation, and a suitable equation of state.  We use \citet{Chabrier2021} for the H/He (with hydrogen and helium mass fractions as $X=0.754$ and $Y=0.246$ respectively), which in turn incorporates the \citet{Militzer2013} computations for the entropy.  For the metals, we use a 50\%/50\% rock/ice mixture by mass from ANEOS \citep{Thompson1990}.  These are placed in a fixed-mass $10~M_\oplus$ core; the remaining metal is mixed into the envelope via the additive volumes approximation\footnote{In cases where the planet is highly enriched in metals, the bulk of the metals are in the more compressible envelope rather than the core, leading to a smaller radius at a given metal enrichment and mass than a planet with a more massive core.}.

In \citet{Thorngren2018}, the thermal evolution was the sum of the intrinsic luminosity \citep[calculated from the atmosphere models of ][]{Fortney2007} and an anomalous heating power $P_a(F)$ which was to be estimated:
\begin{equation}
    \frac{dE}{dt} = P_a(F) - 4\pi R^2 \sigma_b T_\mathrm{int}^4.
\end{equation}
Here, $\sigma_b$ is the Stefan-Boltzmann constant and $T_\mathrm{int}$ is the intrinsic temperature. This model naturally allows the planet to reinflate, as it assumes all anomalous heating is able to efficiently heat the planet's entire interior \citep[e.g.,][]{Komacek2020}.  Conversely, cooling to equilibrium (defined as $dE/dt = 0$) is very fast due to how strongly the intrinsic luminosity responds to changes in the adiabat entropy; this is shown as the dark blue ``deep heating only'' line in Fig. \ref{fig:thermalModel_heating}.  By comparison, standard cooling with no anomalous heating is the light blue ``cool giant'' line, featuring relatively fast cooling and no reinflation.

\begin{figure*}[t]
    \centering
    \includegraphics[width=\linewidth]{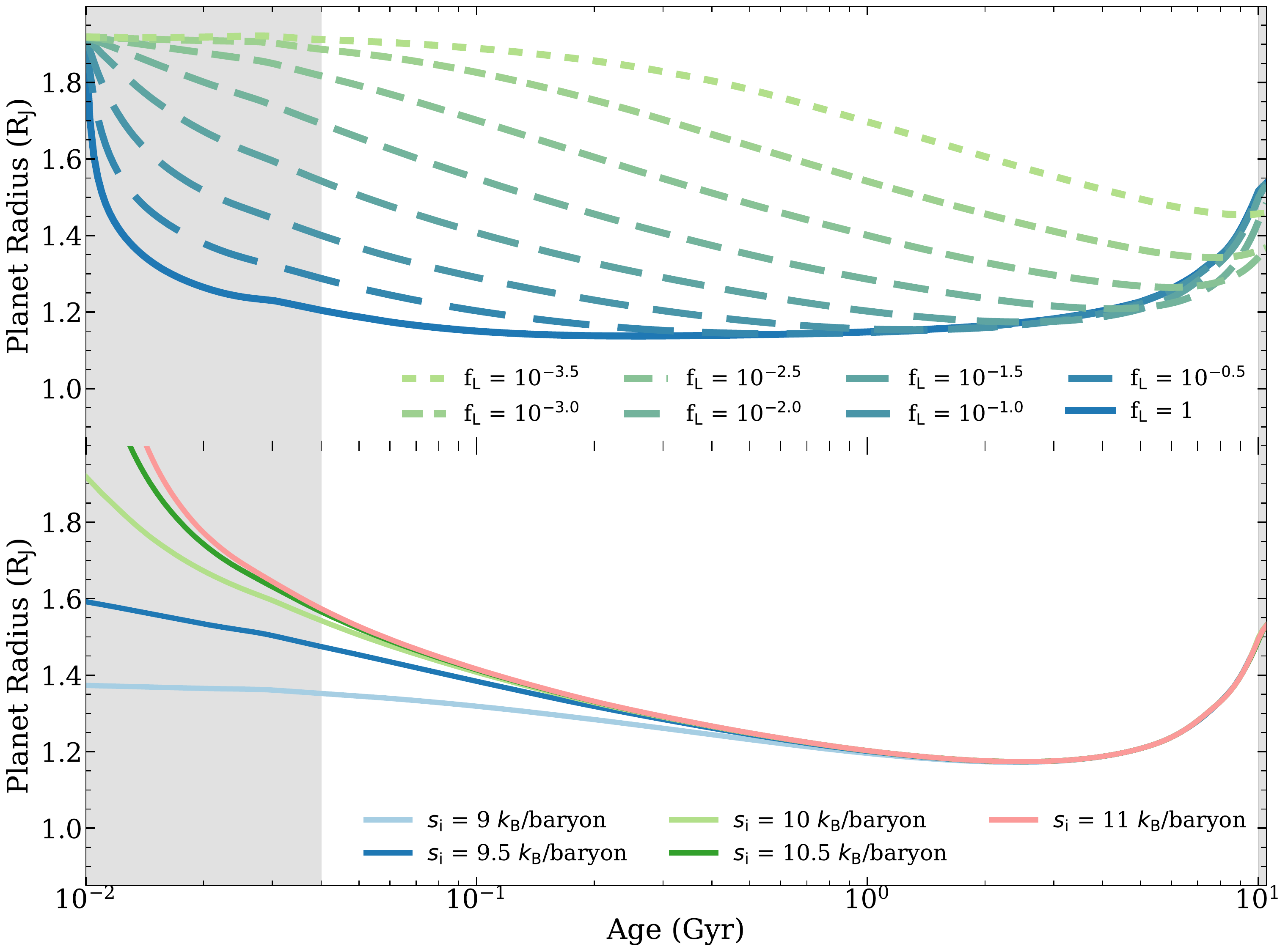}
    \caption{ Demonstration of the effects of differing deep heating fractions or initial specific entropies on a planet's radius evolution. We plot as colored lines the radius of a 1 M$_{\text{Jup}}$ planet with a 10 M$_\oplus$ core and envelope metallicity $Z_{\text{env}}=0.1$ orbiting a 1 M$_\odot$ star at 0.05 AU as a function of time, varying the fraction of total heating caused by deep heating f$_{\text{DH}}$ (assuming $s_{\text{i}} = 10~k_{B}/$baryon) in the upper panel and the initial specific entropy $s_{\text{i}}$ (assuming f$_{\text{DH}} = 10^{-1.5}$) in the lower panel. Larger deep heating fractions cause the planet to reach its equilibrium radius quickly, while smaller deep heating fractions keep it inflated for a longer period of time. 
    At larger values of $s_{\text{i}}$, the planet begins its thermal evolution at a larger radius before cooling over several Gyr. Though the effect of changing $s_{\text{i}}$ vanishes after about 500 Myr, different assumptions for it in our individual retrievals could affect the inferred interior structure of the youngest planets in our sample. 
    Our analysis marginalizes over the hot Jupiter population to extract the population-level aggregate f$_{\text{DH}}$, so we experiment with choosing several values of $s_{\text{i}}$ in our hierarchical Bayesian inference to mitigate any bias due to only exploring one value of it.
    }
    \label{fig:thermalModel_delay}
\end{figure*}

In this work we modify the model to allow for the delayed cooling and reinflation previously inferred \citep[e.g.,][]{Thorngren2021}.  Delayed cooling will be modeled with the intrinsic luminosity reduction factor $f_{\text{L}}$ that directly reduces the intrinsic luminosity of the planet.  If included without any heating term, this modification would produce the light green ``Delayed Cooling'' line in Fig. \ref{fig:thermalModel_heating} -- note the large radii at young ages but a lack of reinflation.  With both delayed cooling and anomalous heating, we have the dark green ``Heated \& Delayed'' line, which features both large but decreasing radii at young ages as well as reinflation once the star ages and brightens sufficiently.

The effects of deep heating and delayed cooling have a compounding effect on the thermal and radius evolution of the planet.  
If $f_{\text{L}}=0.5$, then we would consequently only need half as much deep heating to maintain the $dE/dt=0$ equilibrium discussed in \citet{Thorngren2018} and \citet{Sarkis2020}.  
To guarantee the same equilibrium state as previously observed, we scale both the intrinsic luminosity and the anomalous power by the same factor $f_{\text{L}}$.  Thus our thermal evolution equation becomes
\begin{align}
    \frac{dE}{dt} &= f_{\text{L}} P_a(F) - 4\pi f_{\text{L}} R^2 \sigma_b T_\mathrm{int}^4 + P_\mathrm{rad} \\
    &= \pi R^2 F \epsilon(F)f_{\text{L}}  - 4\pi R^2 \sigma_b T_\mathrm{int}^4 f_{\text{L}} + P_\mathrm{rad}. \label{eq:heating}
\end{align}
Following \citet{Thorngren2023}, we have included radioactive heating $P_\mathrm{rad}$ for completeness; however, it is negligible for hot giants.
We have also rewritten the anomalous heating as a fraction of the total instellation $\epsilon$, which we will refer to as the heating efficiency.  
Fig. \ref{fig:thermalModel_delay} shows the radius evolution of an example planet for various values of $f_{\text{L}}$ and the median heating efficiency given by \citet{Thorngren2018}.  
Only values of $f_{\text{L}}$ very close to zero can produce strong delayed cooling and begin to suppress reinflation.

Although we use an intrinsic luminosity reduction factor to parameterize this model, many hot Jupiter heating models achieve this delay through shallow heating (see Sec. \ref{sec:Introduction}).  Such effects could predominantly act by \emph{replacing} outgoing heat from the interior rather than simply preventing its escape.  In these cases, the intrinsic temperature observed in atmospheric mixing \citep[e.g.,][]{Sing2024} would be the same as if the heating was all deep (no delayed cooling) despite the planet interior cooling more slowly.  On the other hand, it is unclear how this would affect dynamo-dominated magnetic field strengths \citep[see][]{Yadav2017}.  Regardless, it is useful to note that our model is equivalent to shallow- and deep-heating powers of:
\begin{align}
    P_s &= 4\pi R^2\sigma_b T_\mathrm{int}^4 (1-f_{\text{L}})\\
    P_d &= \pi R^2 F \epsilon(F)f_{\text{L}}.
\end{align}
Thus out-of-equilibrium shallow heating depends on $T_\mathrm{int}$ rather than $F$.  This is necessary to achieve delayed cooling while still matching the observed radii -- if we simply partitioned a flux-dependent heating into deep and shallow portions, we \emph{cannot} reproduce both.  While it seems more likely that shallow heating depends on both $T_\mathrm{int}$ and the instellation $F$, such a model would be infeasible to fit from population radii.  The model presented here should therefore be understood as a next-lowest-order approximation to the heating above that of \citet{Thorngren2018}.

\section{Host Star \& Planet Property Inference} \label{sec:stellarInference}

The inference of precise and accurate fundamental stellar parameters requires a precise parallax measurement.
We therefore require Gaia Data Release (DR) 3 \texttt{source\_id} identifiers for each hot Jupiter host in our inference sample\footnote{For the details of Gaia DR3 and its data processing,
see \citet{Gaia_mission2016, GaiaEDR3, GaiaDR3Frame, GaiaDR3Summary},
\citet{GaiaEDR3Validation}, \citet{lin21a, lin21b}, \citet{GaiaEDR3XMatch,
GaiaDR3XMatch}, \citet{EDR3photometry}, \citet{row21}, \citet{tor21}, and
\citet{GaiaDR3Validation}.}.
We select from the NASA Exoplanet Archive's \texttt{pscomppars} table \citep{ExoplanetArchive, Christiansen25, PSCompPars} as of July 22, 2025 all exoplanet systems available at the time with both a radius (\texttt{pl\_radj}) and mass or minimum mass (\texttt{pl\_bmassj}) measurement.
We do not yet set limits on mass or radius as the stellar masses and radii we infer can cause the inferred planet masses and radii to differ from the reported composite parameters from the table.
We assemble a list of Gaia DR3 \texttt{source\_id} identifiers in three steps, following the procedure of \citet{Schmidt26a}.
We first use the Gaia DR2 \texttt{source\_id} identifier provided by the NASA Exoplanet Archive to find the DR3 \texttt{source\_id} using the \texttt{dr2\_neighbourhood} table.  
If a NASA Exoplanet Archive entry lacks a Gaia DR2 \texttt{source\_id} but has a TESS Input Catalog \citep[TIC; ][]{TIC2019} ID, then we use the TIC ID to obtain a Gaia DR2 \texttt{source\_id} and then follow the same procedure.  
If a NASA Exoplanet Archive entry lacks both a Gaia DR2 \texttt{source\_id} and a TIC ID, then we use \texttt{astroquery} \citep{astroquery} to query Simbad for the corresponding Gaia DR3 \texttt{source\_id} corresponding to the system \texttt{hostname}.
This procedure results in a sample of 1512 host stars amenable to our stellar property inference methodology.

We infer the fundamental and photospheric stellar parameters for the giant planet host stars in our sample using the \texttt{isochrones} \citep{mor15} package to execute with \texttt{MultiNest} \citep{fer08,fer09,fer19} a simultaneous Bayesian fit of the Modules for Experiments in Stellar Evolution \citep[MESA;][]{pax11,pax13,pax18,pax19,jer23} Isochrones \& Stellar Tracks \citep[MIST;][]{dot16,cho16} isochrone grid to a curated collection of data for each star. 
Depending on the star, we fit the MIST grid to a subset of the following data:
\begin{enumerate}
\item Galaxy Evolution Explorer \citep[GALEX;][]{mart05} near-ultraviolet (NUV) photometry from the GUVcat\_AIS \citep{bia17}, or, if the system is in the Kepler field, the GALEX-CAUSE Kepler (GCK) survey \citep{olm15} including in quadrature a zero-point uncertainty of 0.02 mag in both cases;
\item SkyMapper Southern Survey Data Release (DR) 4 $uvgriz$ photometry including in quadrature their zero-point uncertainties (0.03, 0.02, 0.01, 0.01, 0.01, 0.02) mag \citep{onk24};
\item the most up-to-date Sloan Digital Sky Survey (SDSS) $ugriz$ photometry from DR13 including in quadrature their zero-point uncertainties (0.015, 0.003, 0.003, 0.003, 0.003) mag \citep{fuk96, gun98, gun06, yor00, doi10, fin16, alb17};
\item Panoramic Survey Telescope \& Rapid Response System (Pan-STARRS) DR2 $grizy$ photometry including in quadrature their zero-point uncertainties (0.014, 0.014, 0.015, 0.015, 0.018) mag \citep{cha16,fle20,mag20a,mag20b,mag20c,wat20};
\item Kepler Input Catalog (KIC) $griz$ photometry (if the system is in the Kepler field) including in quadrature their zero-point uncertainties of about 0.03 mag \citep{bro11};
\item Tycho-2 $B_{T}$ and $V_{T}$ photometry including in quadrature their zero-point uncertainties (0.078, 0.058) mag \citep{hog00,marz05};
\item Gaia EDR3 $G$ photometry including in quadrature its zero-point uncertainty \citep{Gaia_mission2016, GaiaEDR3, GaiaEDR3Validation, EDR3photometry, row21, tor21};
\item Two-micron All-sky Survey (2MASS) $JHK_{s}$ photometry including their zero-point uncertainties \citep{skr06};
\item Wide-field Infrared Survey Explorer (WISE) All-sky \citep{wri10} or AllWISE \citep{mai11} $W1$ and $W2$ photometry in quadrature their zero-point uncertainties (0.032,0.037) mag\footnote{\url{https://wise2.ipac.caltech.edu/docs/release/allsky/expsup/sec4\_4h.html\#PhotometricZP}}.
Because the procedures used in the construction of the WISE All-sky catalog were better for saturated sources than the procedures used in the construction of the AllWISE catalog, we use WISE All-sky data when $2 \leq W1 < 8$ and $1.5 \leq W2 < 7$.  We use AllWISE data when $W1 \geq 8$ and $W2 \geq 7$.
\item We also fit to a zero point-corrected Gaia EDR3 parallax \citep{GaiaEDR3, GaiaEDR3Validation, lin21a, lin21b, row21, tor21} and
\item when available an estimated extinction value based on the \citet{gre19} three-dimensional extinction map calculated using the \texttt{dustmaps} Python module \citep{gre18}.  If an extinction estimate is unavailable from that source, then we use a different three-dimensional extinction map \citep{lal22,ver22}.
\end{enumerate}
We use the data quality flags described in \citet{Hamer22}.  We only include ground-based optical data from one survey in each fit, prioritizing the survey that provides ultraviolet data.  If more than one survey provides ultraviolet data, we then use data from the survey that provides the most independent flux measurements.  If no survey provides ultraviolet data, then we use data from the survey that provides the most independent flux
measurements.

For our priors, we use a \citet{cha03} log-normal mass prior for $M_{\ast} < 1~M_{\odot}$ joined to a \citet{sal55} power-law prior for $M_{\ast} \geq 1~M_{\odot}$, a metallicity prior based on the Geneva-Copenhagen Survey \citep[GCS;][]{cas11}, a log-uniform age prior between 100 Myr and the age of the universe 13.721 Gyr, a uniform extinction prior in the interval between the value described above minus/plus five times its uncertainty, and a distance prior proportional to volume between the \citet{bai21} geometric distance minus/plus five times its uncertainty.

Next, we obtain planet parameters for the exoplanets in systems for which an \texttt{isochrones} stellar property inference was successful.
We select our sample of exoplanet system measurements from the NASA Exoplanet Archive's \texttt{ps} table \citep{ExoplanetArchive, Christiansen25, PS}.
We eliminate systems discovered by pulsar timing, microlensing, and direct imaging as planet radii cannot be directly inferred for these systems.
Our process for calculating planet parameters prefers to use as few Exoplanet Archive entries as possible for each planet's orbit solution (Doppler semiamplitude $K$ and eccentricity $e$) and transit solution (orbital period $P_{\text{orb}}$, inclination $i$ or impact parameter $b$, and planet-to-star radius ratio $q$ or transit depth $\delta$).
We prefer transit solutions with inclinations over impact parameters and radius ratios over transit depths as inclinations and radius ratios do not require additional transformations to obtain a mass or radius inference.
In cases where a transit depth is reported instead of a planet-to-star radius ratio, we transform the transit depth radius ratio by taking its square root and transform its uncertainty in quadrature.
We choose the larger of the upper and lower uncertainty for each parameter in cases where they are unequal as the inputs to our hierarchical Bayesian inference must be normally distributed.

Our process for choosing the best combination of orbit and transit solutions requires calculating planet masses and radii for each combination of measurement entries for a given planet.
We generate posteriors for the orbital period, Doppler semiamplitude, inclination, and eccentricity of the same size as the \texttt{isochrones} stellar mass posterior as normal distributions based on our selected parameter sets' values and uncertainties.
We also generate a posterior for the planet-to-star radius ratio of the same size as the \texttt{isochrones} stellar radius posterior in the same way.
In some circumstances there are no entries for a planet with uncertainties on one or more parameters, or the table misplaces the uncertainty values.
For these we are unable to fold in those measurement uncertainties when calculating the mass and radius for the planet.
We infer masses for our planets by solving the binary mass function
\begin{equation}\label{eq:binarymassfunction}
   f(M) = \frac{M_{2}^3 \sin^3{i}}{\left(M_{1} + M_{2}\right)^2} = \frac{P K_1^3}{2 \pi G} \left(1 - e^2\right)^{3/2}.
\end{equation}
In a similar fashion to \citet{Schmidt23}, we re-parameterize Equation (\ref{eq:binarymassfunction}) as a cubic equation
\begin{equation} \label{eq:morecorrectradialvelocity}
   \sin^3i\left(\frac{M_{2}}{M_{1}}\right)^{3} - B \left(\frac{M_{2}}{M_{1}}\right)^{2} - 2 B\left(\frac{M_{2}}{M_{1}}\right) - B = 0,
\end{equation}
where the coefficient $B$ is defined as
\begin{equation}\label{eq:B}
   B = \frac{1}{2 \pi G} \frac{P K_{1}^{3} \left(1-e^{2} \right)^{3/2}}{M_{1}}.
\end{equation}
Taking the real root of Equation (\ref{eq:morecorrectradialvelocity}) results in a mass ratio that can then be used to determine secondary masses given primary masses. 
We use the Exoplanet Archive parameters and \texttt{isochrones}-generated posteriors with Equation (\ref{eq:morecorrectradialvelocity}) to calculate planet mass and radius posteriors, selecting the combination where the sum of the ratio between the mass and its uncertainty and the ratio between the radius and its uncertainty is highest.
With the inferred masses and radii in hand, we restrict our sample to 408 hot Jupiters with P$_{\text{orb}} < 10$ d and $0.1 < M_{p} < 10$ M$_{\text{Jup}}$ using the orbital period and mass lower limit advocated by \citet{Wright12} and the giant planet mass upper limit of 10 M$_{\text{Jup}}$ proposed by \citet{Schlaufman2018}.

\section{Hierarchical Bayesian Inference} \label{sec:hierarchicalBayesianInference}
Our two-level hierarchical Bayesian model for $N$ planets contains $6N+2$ parameters: $\mathbf{M}$, $\mathbf{Z}$, $\mathbf{M_*}$, $\mathbf{\mathrm{[Fe/H]}}$, $\mathbf{t}$, $\mathbf{h}$, $f_{\text{L}}$, $H$, where boldface indicates that the quantity is a vector with one element for each planet in our sample. 
We show the definitions of these variables in Table \ref{tab:bayesian}. The semimajor axis $a^i$ and eccentricity $e^i$ of each planet's orbit are considered fixed and so are not mentioned in the probability distributions.
For the sake of organization and brevity, we define the vector $k^i$ as the vector of planet parameters for the $i^\mathrm{th}$ planet, excluding the heating model parameter $h^i$:
$\ki = \{M^i, Z^i, M_*^i, \mathrm{[Fe/H]}^i, t^i\}$.  We define $\mathbf{K}$ (capitalized without a subscript) to be the $N\times5$ matrix formed by stacking each $k_i$ as a row. 
Similarly, $\mathbf{h}$ is the vector formed by stacking $h^i$. 
All equations that follow assume the parameters of the observational distributions are given.

\begin{deluxetable}{ll}[t]
    \tablecaption{Bayesian Model Variables}
    \label{tab:bayesian}
    \tablehead{\multicolumn{2}{c}{\textbf{Model Parameters}}}
    \startdata
    $M^i$, $\mathbf{M}$ & Planetary true mass\\
    $Z^i$, $\mathbf{Z}$ & Bulk planet metallicity (by mass)\\
    $h^i$, $\mathbf{h}$ & Heat factor (see Section \ref{sec:hierarchicalBayesianInference}) \\
    $M_*^i$ & Parent star true mass \\
    $\text{[Fe/H]}^i_0$ & Parent star zero-age true metallicity \\
    $t^i$ & Parent star true age. \\
    \hline
    \multicolumn{2}{c}{ \textbf{Hyperparameters}} \\
    \hline
    $f_{\text{L}}$ & Intrinsic luminosity reduction factor \\
    $H$ & Population mean heat factor \\
    \hline 
    \multicolumn{2}{c}{ \textbf{Constants}} \\\hline
    $M_\mathrm{obs}^i$, $\sigma_M^i$ & Observed planet mass and uncertainty \\
    $R_\mathrm{obs}^i$, $\sigma_R^i$ & Observed planet radius and uncertainty \\
    $M_\mathrm{*obs}^i$ & \texttt{isochrones} stellar mass \\
    $\text{[Fe/H]}^i$ & \texttt{isochrones} stellar photospheric metallicity \\
    $t_\mathrm{obs}^i$ & \texttt{isochrones} stellar age \\
    $a_\mathrm{obs}^i$ & semimajor axis, fixed \\
    $e_\mathrm{obs}^i$ & eccentricity, fixed \\
    \hline 
    \multicolumn{2}{c}{ \textbf{Agglomerations}} \\\hline 
    $\ki$, \textbf{K} & $\{M^i, Z^i, M_*^i, \mathrm{[Fe/H]}^i_0, t^i\}$\\
    $\mathbf{\mu_\mathrm{obs}^i}$ & \{$M_\mathrm{*obs}^i$, $\text{[Fe/H]}^i$, $t_\mathrm{obs}^i$\}\\
    $\mathbf{\Sigma_*^i}$ & Covariance($M_\mathrm{*obs}^i$, $\text{[Fe/H]}^i$, $t_\mathrm{obs}^i$)
    \enddata
    \tablecomments{The superscript $i$ indicates the $i^\mathrm{th}$ planet in our sample, and bolded variables are the vector of that parameter for all planets.  The covariance $\mathbf{\Sigma_*^i}$ is output by the \texttt{isochrones} run for each host star. Additionally, as the MIST grids used by \texttt{isochrones} lack the initial bulk metallicity column, we fit for this using the stellar age, mass, and present-day photospheric metallicity to obtain the initial bulk metallicity for each system (see Section \ref{sec:hierarchicalBayesianInference:sampling} for more details).}
\end{deluxetable}

To handle possible variations in the heating efficiency $\epsilon(F)$ (Eq. \ref{eq:heating}), we parameterize the variation with $h^i$, which is the offset relative to the mean heating in number of standard deviations. 
To match the results of \citet{Thorngren2018} exactly, we could set the prior as $h\sim\mathcal{N}(0, 1^2)$ where $\mathcal{N}(\mu, \sigma^2)$ is a normal distribution with mean $\mu$ and standard deviation $\sigma$. 
However, to allow the model additional flexibility while maintaining the functional form of $\epsilon(F)$ we instead allow a global mean offset to the heating $H$, such that $h^i\sim\mathcal{N}(H,1^2)$.
Due to some differences in equation of state between this article and \citet{Thorngren2018}, we expect $H$ to be somewhat negative (i.e. less heating is required).

The likelihood for the $i^\mathrm{th}$ planet is:
\begin{align}
    p(\robs^i | \ki, h^i, f_{\text{L}}) = \mathcal{N}\left(
        \robs^i | R(\ki, h^i, f_{\text{L}}), {\sigma_R^i}^2
    \right). \label{eq:individualPrior}
\end{align}

The posterior for a single planet is therefore:
\begin{align}
    p(\ki, h^i, f_{\text{L}}| R_\mathrm{obs}) \propto p(R_{obs}^i | \ki, h^i, f_{\text{L}}) p(\ki) p(f_{\text{L}}) p(h^i).
\end{align}

The prior for $\ki$ is the product of the priors for its elements:
\begin{align}
    & p(\ki) = p(M^i) p(Z_i|M) p(M_*^i, \mathrm{[Fe/H]}^i, t^i) \label{eq:kPrior}\\
    & p(M^i) \sim \mathcal{N}(M_\mathrm{obs}^i, {\sigma_M^i}^2) \\
    & p(M_*^i, \mathrm{[Fe/H]}^i, t^i) = \mathcal{N}_3(\mathbf{\mu^i_*}, \mathbf{\Sigma_*^i}) \\
    & p(h^i) \sim \mathcal{N}(H, 1^2).
\end{align}

Here $\mathcal{N}_3$ is a multivariate normal distribution with 3 dimensions, $\mu_*^i$ is the vector means of \textit{observed} stellar parameters $\mathbf{M_*}$, $\mathbf{\mathrm{[Fe/H]}}$, and $\mathbf{t}$, and $\mathbf{\Sigma_*^i}$ is their covariance matrix. 
Combining the individual planet posterior distributions, we have the overall posterior
\begin{align}
    p(\mathbf{K},\textbf{h},f_{\text{L}},H) &\propto p(f_{\text{L}}) p(H) \prod_{i=1}^N p(R_{obs}^i | \ki, h^i, f_{\text{L}}) p(\ki) p(h^i).
\end{align}

The terms within the product are just equations \ref{eq:individualPrior} and \ref{eq:kPrior}.  We set our hyperpriors for $f_{\text{L}}$ and $H$ as:
\begin{align}
    \log(f_{\text{L}})& \sim \mathcal{U}(-4, 0) \\
    H &\sim \mathcal{N}(0, 1^2)
\end{align}
where $\mathcal{U}(a,b)$ is a uniform distribution between $a$ and $b$.

\subsection{Sampling Strategy}\label{sec:hierarchicalBayesianInference:sampling}
While the previous subsection contains a complete statistical model description, the number of parameters is so high that sampling directly from it is impractical. 
As we are only interested in the hyperparameter $f_{\text{L}}$, we can marginalize out the individual parameters and thereby break the sampling process into more manageable chunks. 
The effective sampling rate of the Metropolis-Hastings algorithm scales as the square of the number of parameters, so it vastly reduces the required computation time to first run a Markov Chain Monte Carlo simulation (MCMC) for each planet individually and then run a final MCMC to combine the results into a posterior for $f_{\text{L}}$.

We begin by defining an auxiliary function $Q^i$,
\begin{align}
    Q^i(f_{\text{L}}, h^i) &:= \int p(R_{obs}^i | \ki, f_{\text{L}}) p(\ki) \mathcal{N}(h^i|0, 2^2)\mathrm{d}\ki.
\end{align}
such that there is one $Q^i$ function per planet.
These are calculated by sampling from the individual planet posteriors and ignoring the posterior parameter samples other than $f_{\text{L}}$ and $h^i$ when computing the integral.
The function is then reconstructed using a 2-dimensional Gaussian kernel density estimation (KDE).
We then rewrite the upper level of the hierarchical prior, marginalized over $\ki$, in terms of these $Q$ functions:
\begin{align}
    p(f_{\text{L}},H) &\propto p(f_{\text{L}}) p(H) \prod_{i=1}^N Q^i(f_{\text{L}}) \frac{p(h^i|H)}{\mathcal{N}(h^i|0, 2^2)}. \label{eq:finalPosterior}
\end{align}
Note that we did not know the value of $H$ while running the individual planet MCMCs.
As such, we instead imposed a prior on $h^i$ where $p(h^i)\sim\mathcal{N}(0,2^2)$ and then added a correction factor to each planet in equation \ref{eq:finalPosterior}.
This strategy amounts to importance sampling and is an excellent approximation so long as $|H|$ isn't much more than 1 and an adequate number of samples are taken.

\begin{deluxetable*}{lcccccccc}
    \centering
    \tablecaption{Hierarchical Bayesian Inference Hot Jupiter Sample}
    \label{tab:sample}
    \tablehead{Planet Name & $M^i_{\text{obs}}$ (M$_{\text{Jup}}$) & $R^i_{\text{obs}}$ (R$_{\text{Jup}}$) & $e$ & $P_{\text{orb}}$ (d) & $M^i_{\ast \text{obs}}$ (M$_\odot$) & [Fe/H]$^i$ & $t^i_{\text{obs}}$ (Gyr) & $R_\ast$ (R$_\odot$)}
    \startdata
    CoRoT-1\,b & $1.12_{-0.08}^{+0.08}$ & $1.66_{-0.04}^{+0.03}$ & 0 & 1.508977 & $1.07_{-0.05}^{+0.11}$ & $-0.14_{-0.06}^{+0.08}$ & $4.8_{-3.0}^{+1.6}$ & $1.24_{-0.02}^{+0.02}$  \\
    CoRoT-12\,b & $0.85_{-0.06}^{+0.06}$  & $1.30_{-0.04}^{+0.05}$  & 0.05  & 2.828042 & $0.95_{-0.05}^{+0.08}$  & $0.02_{-0.07}^{+0.09}$  & $7.5_{-4.9}^{+3.4}$ & $1.01_{-0.03}^{+0.03}$  \\
    CoRoT-13\,b & $1.22_{-0.07}^{+0.07}$  & $1.06_{-0.04}^{+0.04}$  & 0 & 4.03519 & $0.99_{-0.06}^{+0.05}$  & $-0.15_{-0.08}^{+0.08}$  & $7.6_{-1.9}^{+2.7}$ & $1.19_{-0.04}^{+0.04}$  \\
    CoRoT-18\,b & $3.47_{-0.15}^{+0.13}$  & $1.18_{-0.03}^{+0.03}$  & 0 & 1.90009 & $0.95_{-0.06}^{+0.04}$  & $-0.01_{-0.07}^{+0.05}$  & $3.4_{-2.3}^{+4.7}$ & $0.91_{-0.02}^{+0.02}$ \\
    CoRoT-19\,b & $1.07_{-0.06}^{+0.07}$  & $1.25_{-0.02}^{+0.02}$  & 0.047 & 3.89713 & $1.14_{-0.05}^{+0.10}$ & $-0.11_{-0.06}^{+0.05}$  & $5.6_{-1.8}^{+1.1}$ & $1.63_{-0.02}^{+0.02}$  \\
    \enddata
    \tablecomments{This table is sorted by planet name and is available in its entirety in machine-readable format. The sample enumerated by this table has not been vetted for population-level biases and should not be used for occurrence analyses.}
\end{deluxetable*}

Our thermal evolution model's stellar model grid is adapted from the MIST grid and takes the star's initial bulk metallicity as an input.
However, the MIST model grid used by the \texttt{isochrones} Python package is restricted to a subset of parameters that does not include it; as a result the posteriors instead report the present-day photospheric metallicity.
This requires us to use a root finder to calculate the initial bulk metallicity posterior for each star from its inferred present-day photospheric metallicity, stellar mass, and stellar age posteriors.
While this process works with no issue for the vast majority of the host stars in our sample, some metal-rich stars lack suitable MIST models at a high enough initial bulk metallicity to match the inferred present-day photospheric metallicity from their \texttt{isochrones} posterior.
Performing the initial bulk metallicity inference procedure cuts off a significant portion of these stars' metallicity posteriors.
We therefore exclude 33 stars for which this occurs to more than a few percent of samples. 

We perform individual interior structure retrievals on all of the planets in our sample for which our model is suitable: 0.1~M$_{\text{Jup}} < M^i_{\text{obs}}<4~$M$_{\text{Jup}}$, 0.5 M$_{\odot} < M^i_{\ast \text{obs}}<2~$M$_{\odot}$, and $-0.75 < \text{[Fe/H]}^i_{\text{0}}<0.5$, totaling 273.
Though we compute our model for each iteration, underlying it is a static model grid with the following parameter ranges and resolutions:
\begin{enumerate}
    \item $M^i$, with 114 steps between 1.0050608060184631 and 9534.844001949032 M$_\oplus$;
    \item $Z^i$, with 21 steps between 0 and 0.9999;
    \item $f_{\text{core}}$, the fraction of mass in the core, with 21 steps between 0 and 1; and
    \item $s_i$, with 225 steps between 5 and 14.
\end{enumerate}
These resulting posterior distributions are often non-Gaussian, so to facilitate convergence we utilize a two-step MCMC process.
We first run a shorter 20,000-iteration initial retrieval with a 5000-iteration burn-in to characterize the covariance, and then run a second 250,000-iteration retrieval with a 5000-iteration burn-in and using the covariance of
the initial run to set the proposal covariance.
We apply a thinning factor of 10 to both of these steps, where we only record a sample every 10 iterations.
In some cases the model fails to converge or otherwise fails to provide plausible results; we exclude these planets from the upper level of the hierarchical model.
We list the planets in our sample for which we successfully retrieve the interior structure following this methodology in Table \ref{tab:sample}.

To conduct our the upper level of the hierarchical model, we first convert the individual planet retrieval posteriors into probability densities using a Gaussian KDE.
These functions are smooth and well approximated by an interpolator over a moderately dense grid.
We therefore sample the KDE in a 100-by-100 grid evenly spaced in $h^i$ and $\log f_{\text{L}}$ space and interpolate over this grid to evaluate the the upper level of the hierarchical model's likelihood for each planet.
This speeds up our sampling across hundreds of planet posteriors by several orders of magnitude without any noticeable loss of accuracy.
We follow the same two-step sampling procedure as with our individual planet retrievals, by this time with 100,000 iterations and a 10,000 iteration burn-in in both our initial and covariance-adjusted sampling runs.

\section{Results and Discussion} \label{sec:Discussion}
\begin{figure*}
    \centering
    \includegraphics[width=\linewidth]{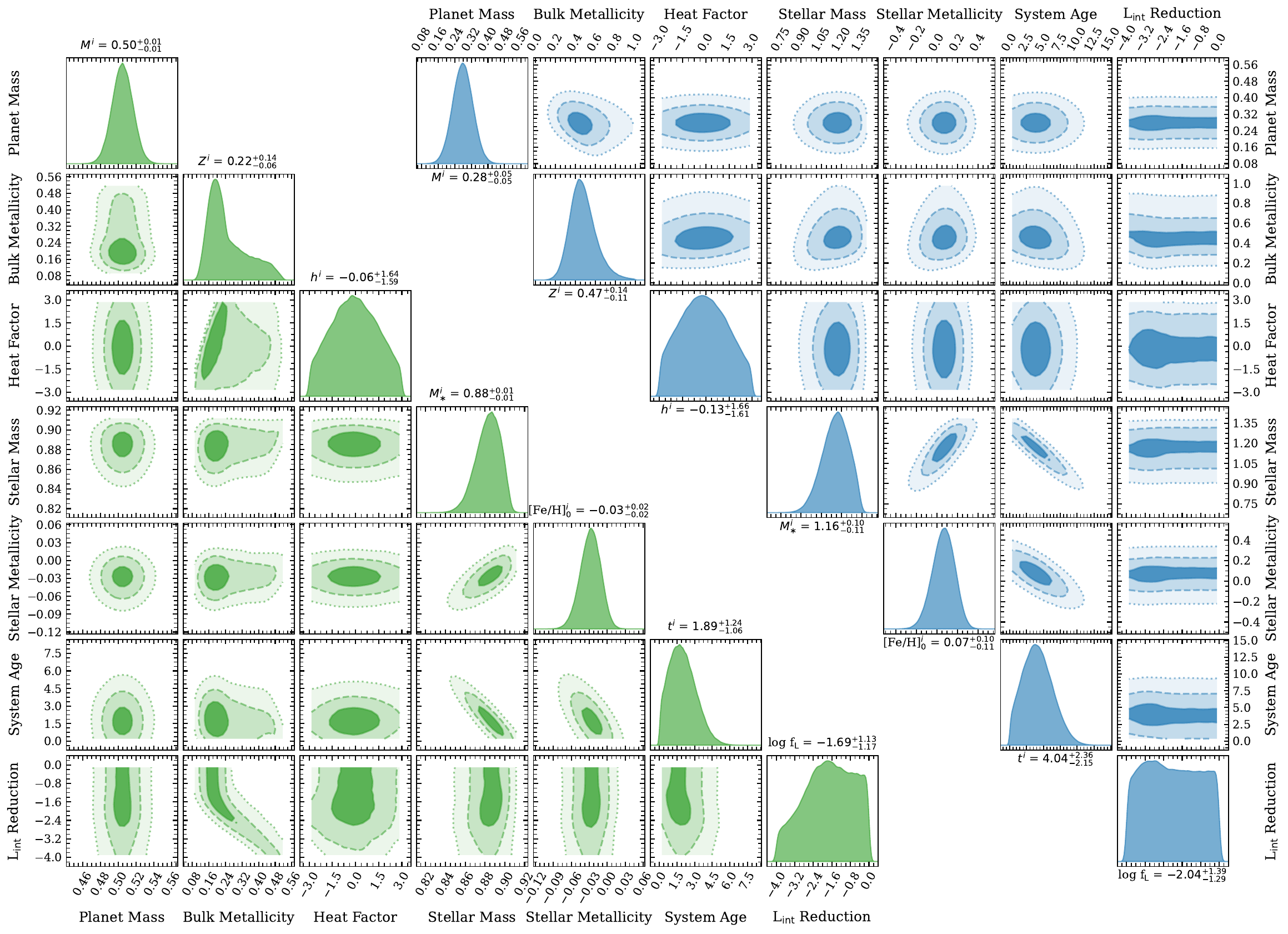}
    \caption{Example interior structure retrieval corner plot for WASP-6\,b (left, green) and CoRoT-25\,b (right, blue), assuming an initial specific entropy of 10 k$_B$ / baryon. The parameters we fit and use as inputs in our model are described in detail in Table \ref{tab:bayesian}. Due to degeneracies between $Z^i$ and f$_{\text{L}}$, it is impossible to infer f$_{\text{L}}$ using only a single planet. This degeneracy is only present for younger planets like WASP-6\,b, meaning that older planets like CoRoT-25\,b are instead useful for constraining differences caused by the use of the updated equation of state.}
    \label{fig:individualPosteriorExample}
\end{figure*}

We show an example corner plot from retrievals on WASP-6\,b (green) and CoRoT-25\,b (blue) in Figure \ref{fig:individualPosteriorExample}.
Our model allows a relatively young planet such as WASP-6\,b to be larger due to either delayed cooling or a less dense envelope caused by a lower bulk metallicity.
This manifests in the posterior as a degeneracy between the planet's bulk metallicity $Z_p$ and intrinsic luminosity reduction factor $f_{\text{L}}$.
On the other hand, a planet like CoRoT-25\,b (Fig. \ref{fig:individualPosteriorExample}, right) is too old to be affected by delayed cooling, having already reached thermal equilibrium where its size is coupled to its host star's evolution (see Sec. \ref{sec:thermalEvolutionModel}).
Such planets are immune to the $Z_p$-$f_{\text{L}}$ degeneracy, making them a natural control group for delayed cooling and useful for constraining the population mean heat factor $H$.
These examples show that a single planet cannot provide enough information to fully constrain $f_{\text{L}}$, making a population-level analysis necessary.

\begin{figure*}
    \centering
    \includegraphics[width=.38\linewidth]{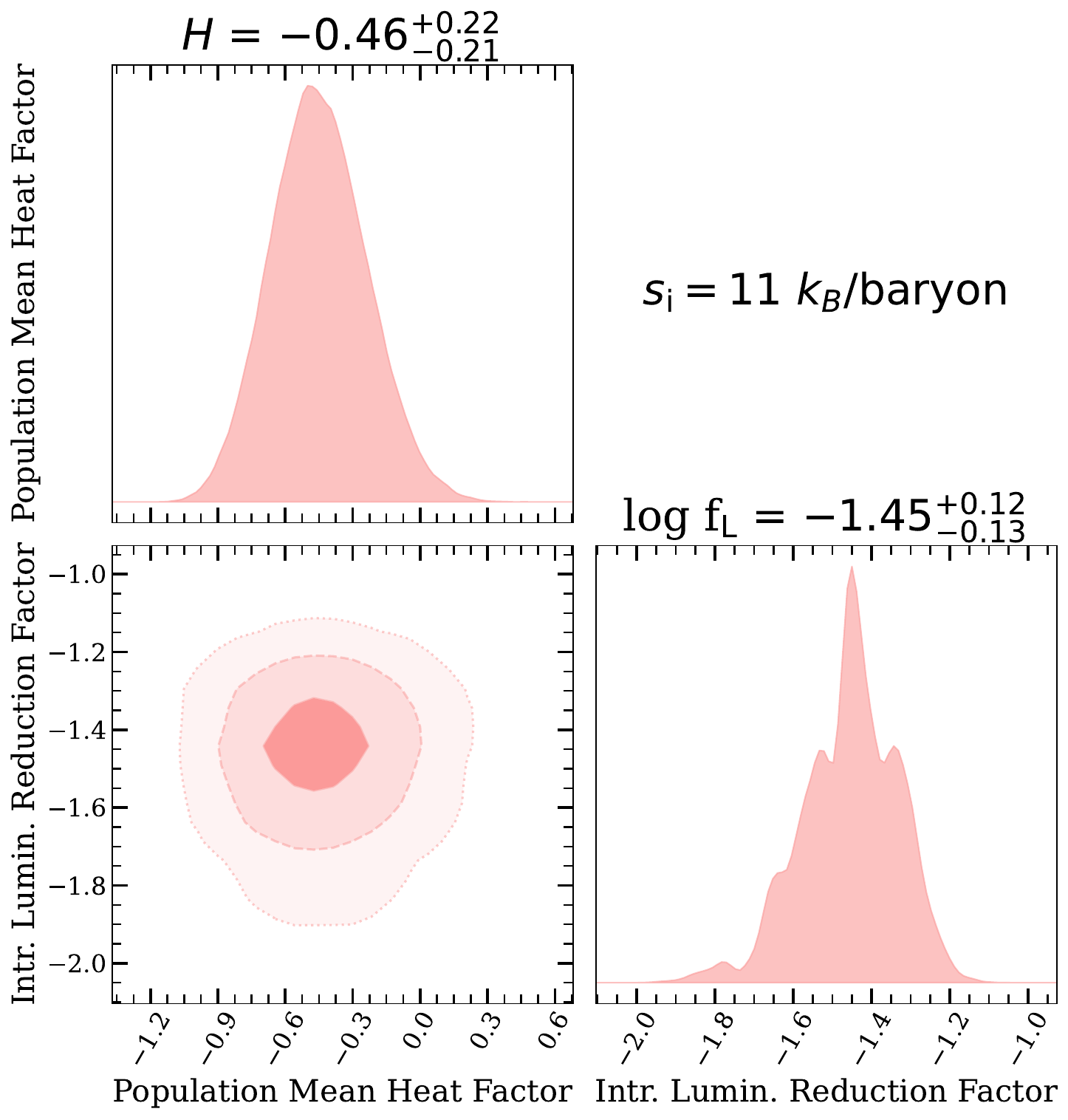}
    \includegraphics[width=.38\linewidth]{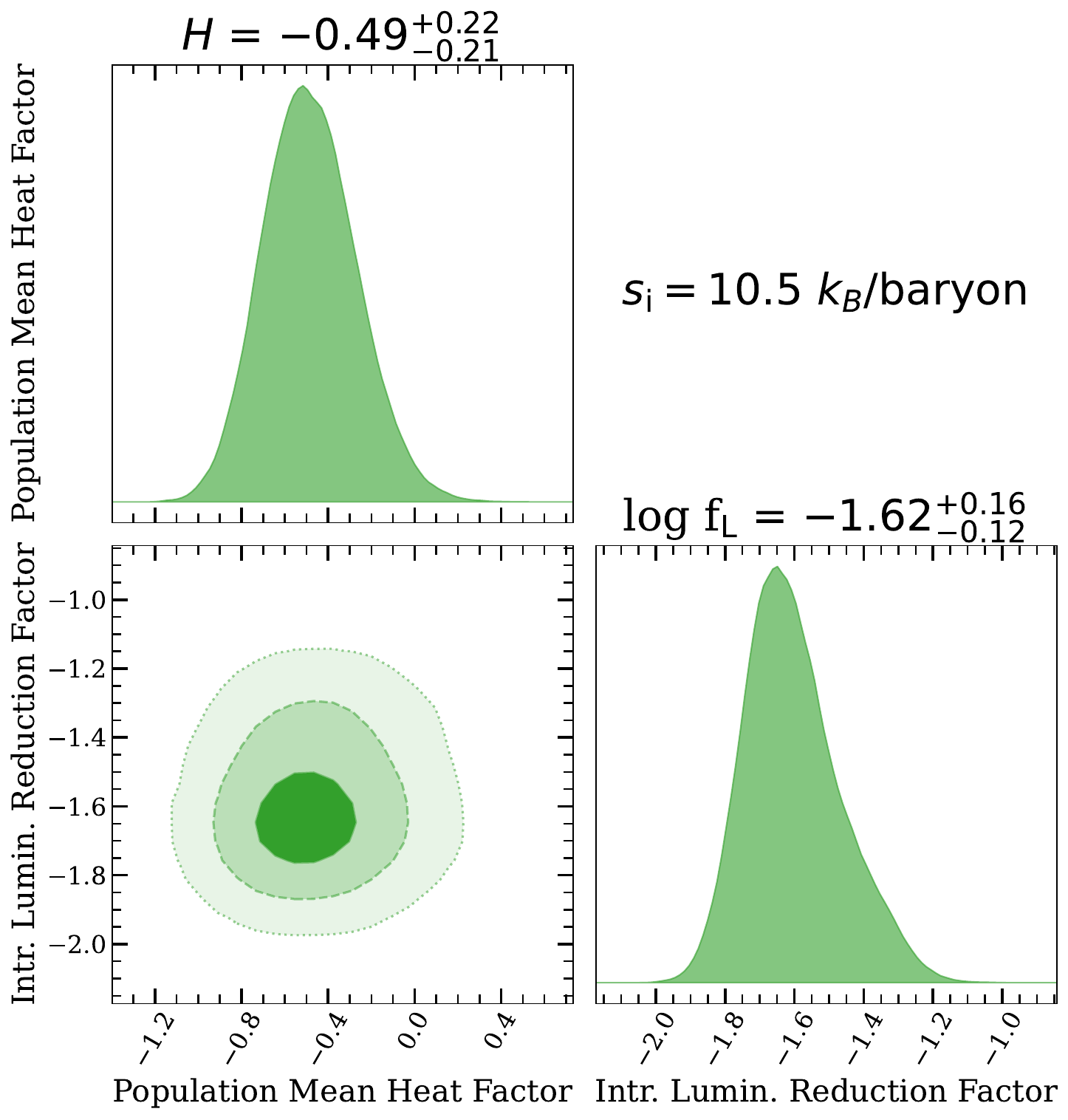}
    \includegraphics[width=.38\linewidth]{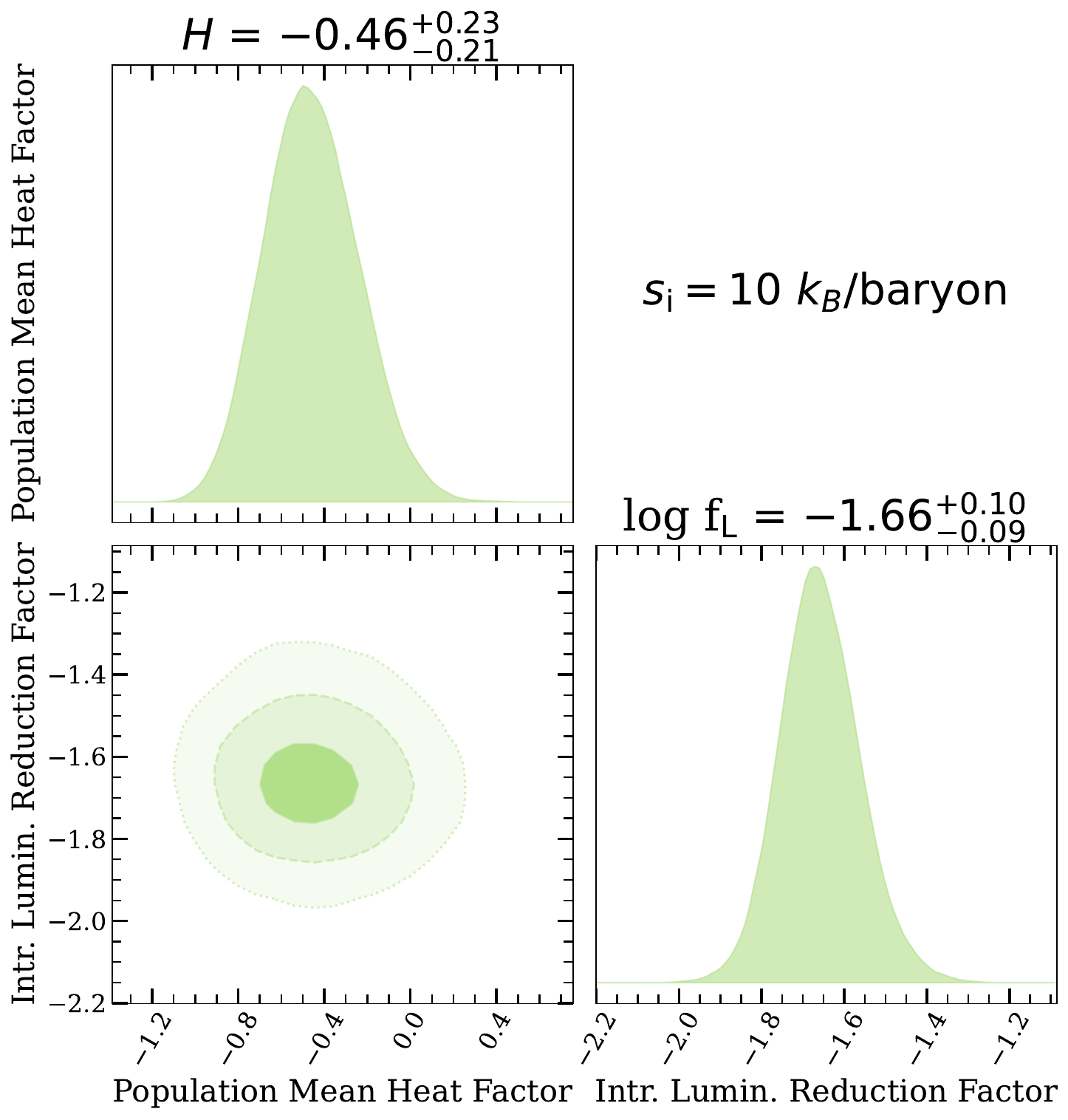}
    \includegraphics[width=.38\linewidth]{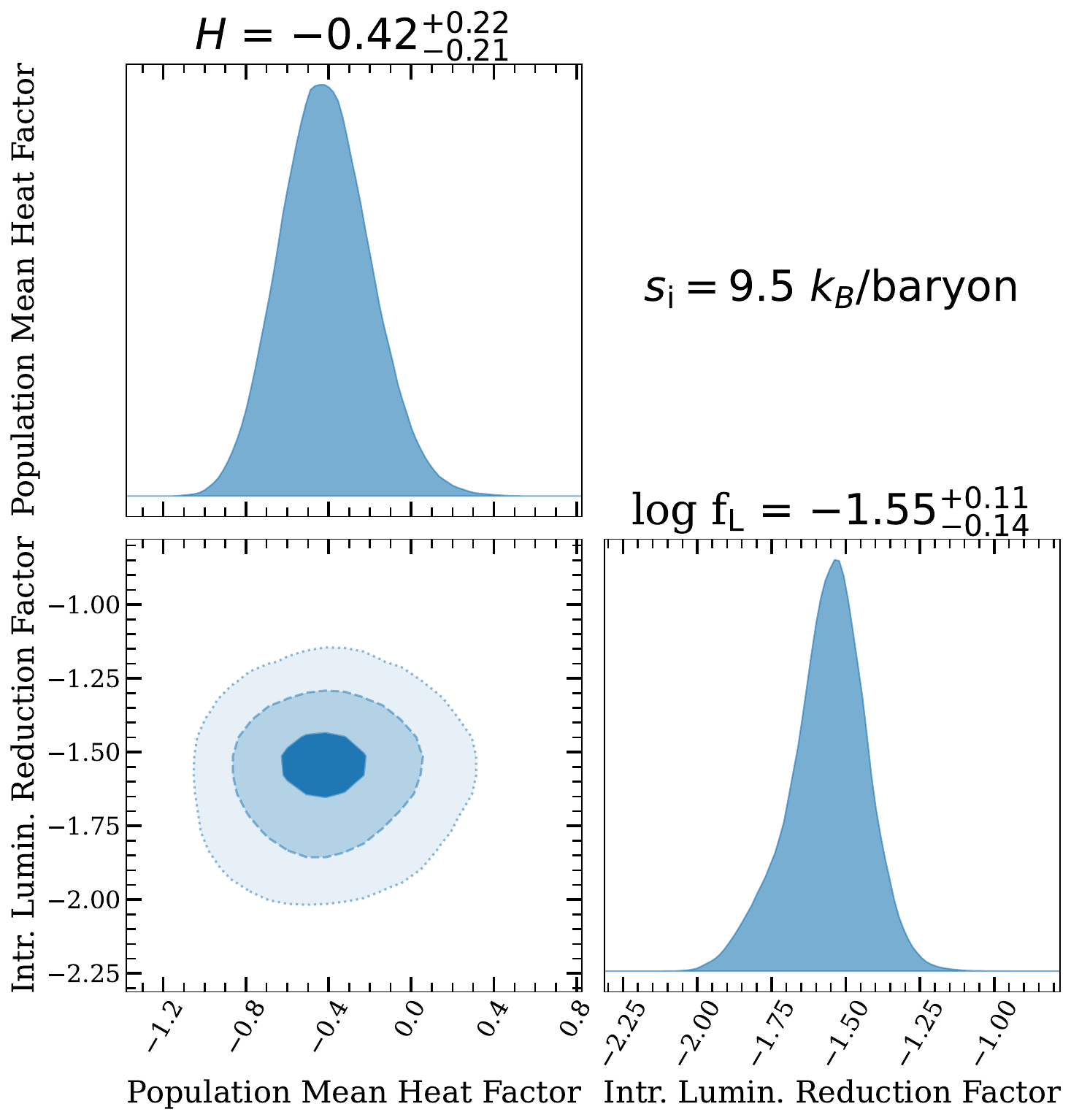}
    \includegraphics[width=.38\linewidth]{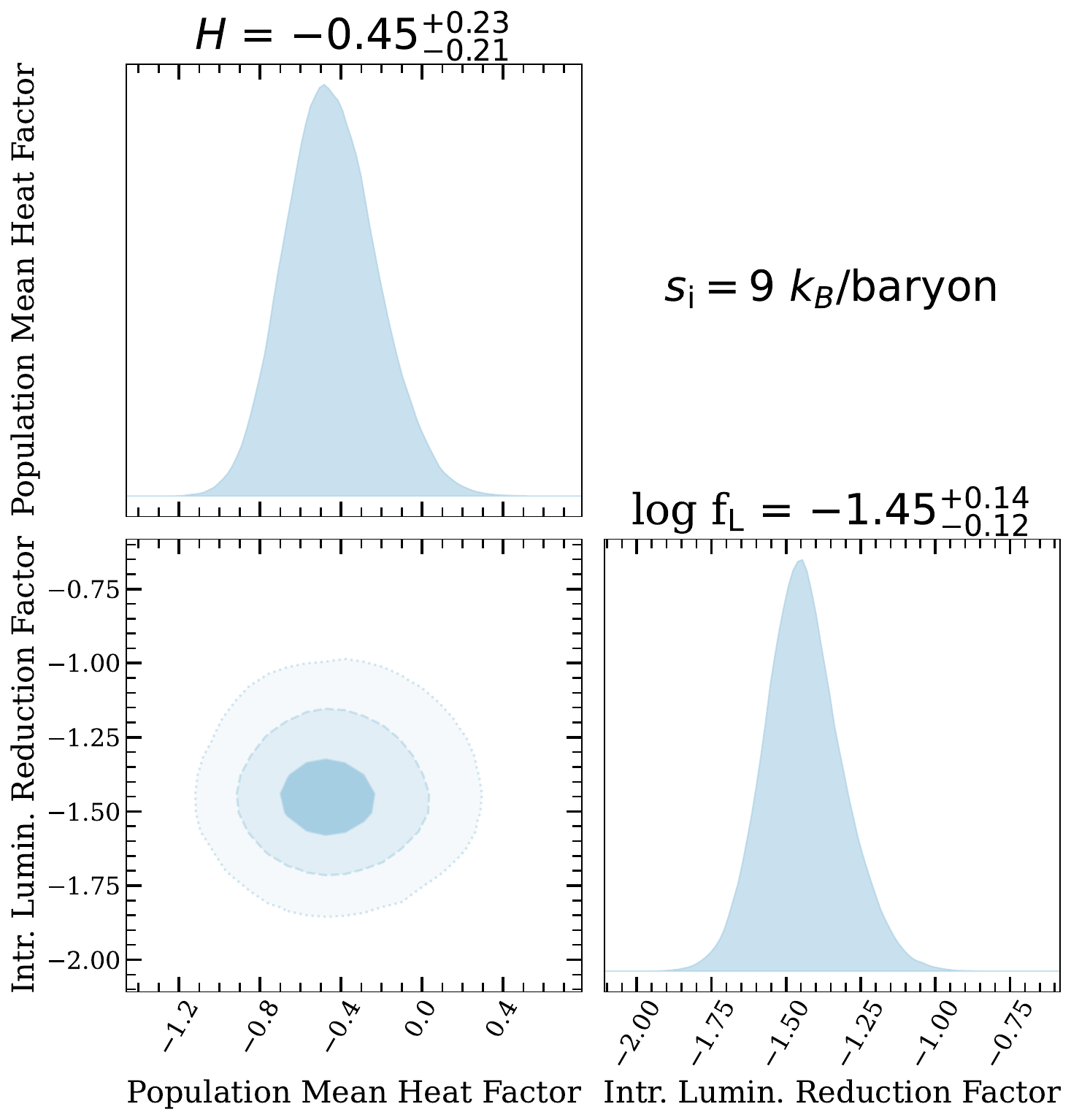}
    \caption{Corner plots for the hyperparameters of our population-level intrinsic luminosity reduction factor $f_{\text{L}}$ and population mean heat factor $H$. Each posterior's model assumes a different initial specific entropy for the individual interior structure retrievals it marginalizes over. Not shown are the individual planets' heat factor hyperparameters, which are included to account for small differences in the theoretical equation of state used between this work and \citet{Thorngren2018}. We find that the interior cooling rates for hot Jupiters are reduced by a factor of 20--60 relative to \citet{Thorngren2018}, regardless of our assumption for the initial specific entropy.}
    \label{fig:finalPosteriors}
\end{figure*}

Next, we show the results of our hierarchical Bayesian inferences in Figure \ref{fig:finalPosteriors}.
Each corner plot in Figure \ref{fig:finalPosteriors} shows the posterior and covariance for our population mean heat factor $H$ and intrinsic luminosity reduction factor $f_{\text{L}}$.
These parameters are uncorrelated, indicating that the delayed cooling effect we see is not biased by or an artifact of our thermal evolution model's equation of state choice.
We vary the value of each retrieved planet's initial specific entropy from $s_{\text{init}}=11~k_B/\text{baryon}$ to $s_{\text{init}}=9~k_B/\text{baryon}$ in increments of $0.5~k_B/\text{baryon}$.
These represent the range between ``hot start'' and ``cold start'' giant planet formation models \citep{Marley2007} to avoid limiting our modeling to a particular assumption about how the planet initially formed.
We note that the shape of the posterior for the $s_{\text{init}}=11~k_B/\text{baryon}$ case is more jagged in comparison the other posteriors; we attribute this to the greater influence of the youngest planets in our sample on it, with their samples being preferred more strongly compared to the posteriors with the lower values of $s_{\text{init}}$.
Regardless, our results are consistent with one another for each value of $s_{\text{init}}$.

We find that the interior cooling rates of the hot Jupiter population are reduced by a factor of between 20 and 60 relative to the simpler heating model presented in \citet{Thorngren2018}.
In cases where the planet has reached thermal equilibrium, this is equivalent to shallow heating accounting for about 97\% of the total heating experienced by hot Jupiters\footnote{Note that while most of the planets in our sample are not exactly in equilibrium, the older ones are not far off from it.}.
In other words, the delayed cooling we observe in the hot Jupiter population means that the youngest ($t^i < $1 Gyr) planets in our sample are consistently larger than the trend inferred by \citet{Thorngren2018}.
We additionally find that the population mean heat $H$ (see Sec. \ref{sec:hierarchicalBayesianInference}) is modest but nonzero to correct for the updated equation of state.
This means that the base amount of heating required to inflate hot Jupiters' radii is slightly lower than \citet{Thorngren2018}\footnote{Also note that, due to the 3D nature of hot Jupiters, the amount of heating we infer here is likely an underestimate \citep{Zhang23}.}. 
The lack of correlation between $H$ and $f_{\text{L}}$ in our posteriors (Fig. \ref{fig:finalPosteriors}) shows that this was not a bias on the latter.

The high intrinsic luminosity reduction factor we find is best explained by predominantly shallow heating (near the surface, rather than the deep interior).
The proposed hot Jupiter heating mechanisms that would cause shallow heating, including Ohmic dissipation \citep{Batygin2010, Batygin2011}, turbulent mixing \citep{Youdin2010}, and longitudinal advection \citep{Tremblin2017} preferentially occur near the radiative--convective boundary (RCB).
The exact location of this heating with respect to the RCB is only of minor importance; while shallow heating mechanisms that operate above it can delay cooling by shifting it deeper into the interior \citep[e.g.,][]{Youdin2010}, very shallow heating cannot cause reinflation on main sequence or post-main sequence timescales \citep{Komacek2020} and is unable to significantly delay cooling \citep{Spiegel13}.
While a larger-than-expected opacity in the shallow interior \citep{Burrows2003, Burrows07} may also contribute, the extent to which our hot Jupiters' radii are inflated at the population level indicates that this explanation alone is insufficient.
Similarly, strong compositional gradients \citep[e.g.,][]{Chabrier2007, Leconte2012} would be insufficient, as neither they nor enhanced opacities could explain the flux-heating relation of \citet{Thorngren2018}.  
As a result, while we cannot definitively rule out these alternatives entirely, they can only be minor contributions compared to shallow heating mechanisms.

Our results strongly exclude $f_{\text{L}}=1$, meaning some delay is required to explain the larger radii of young hot Jupiters. 
On the other hand, exclusively delayed cooling effects from shallow heating would not reinflate planets as seen in hot Jupiters orbiting main sequence \citep{Hartman2016} and evolved \citep{Grunblatt2016, Grunblatt2017} stars.  
Our model's preferred values of $f_{\text{L}}$, a few percent of standard cooling rates, is able to reinflate hot Jupiters with over an order of magnitude less deep heating power than the amount predicted by simpler hot Jupiter anomalous heating models.
This balances the need for both inflated young hot Jupiters and reinflation of older ones, resolving the tension between the deep heating models of \citet{Thorngren2018} and the delayed cooling seen in \citet{Thorngren2021}.
It may also ease the tension between deep heating models and the observed population \citep[e.g.,][in the case of tidal dissipation]{Leconte2010, Yee25}, as much less heating of this kind is necessary to produce the same effect.

Dominant shallow heating has implications for the relationship between circulation-related exoplanet observables and radius inflation.  
This is because the strength of many shallow heating mechanisms is coupled to wind speeds in the atmosphere \citep[see][]{Menou2012}, so circulation-related observables should mirror the heating efficiency-equilibrium temperature trend found in \citet{Thorngren2018}.
Specifically, we expect hot spot offsets for hot Jupiters to scale with the log of the incident flux as $\exp{\left[ (\log(F)-\log(F_0))^2/(2w) \right]}$, where $F_0\approx1.38$ Gerg s$^{-1}$ cm$^{-2}$ and $w\approx0.37$ \citep{Thorngren2024}.
This corresponds to a peak hot spot offset at  $T_\mathrm{eq}\approx1600$ K, decreasing again at higher temperatures.
However, as phase curve offsets are also impacted by the propagation of planetary-scale Kelvin and Rossby waves independent of wind speeds \citep{Lewis22}, confirmation of this relationship will require a large body of high-quality phase curve data to constrain wind speeds.
Though some population studies of archival Spitzer phase curves exist \citep[e.g.,][]{May22, Dang25}, they are limited in sample size as well as the range of equilibrium temperatures and instellations they probe.
Future observing campaigns, as well as missions like European Space Agency's Ariel mission \citep{Tinetti22}, could observe a larger statistical sample to evaluate these predictions.
Other circulation-related observables, such as bulk wind speeds inferred from high-precision Doppler spectroscopy of hot Jupiter atmospheres \citep{Seidel25}, should also exhibit a similar trend as found by \citet{Thorngren2018}.

While our sample was large enough to obtain definitive results, a larger sample of younger ($t^i_{\text{obs}} \lesssim 1$~Gyr) planets with precise masses, radii, and host star properties would enable testing of more complex thermal evolution models.
As hot Jupiter systems experience a range of formation and evolutionary histories, the value of $f_{\text{L}}$ for a given hot Jupiter may not always be constant or equal for both the heating and cooling terms.
The careful study of higher-order effects due to tidal circularization and inspiral \citep[e.g.,][]{Jackson2008,Leconte2010} and/or late-arrival scenarios \citep[e.g.,][]{Hamer22, Schmidt26a} further altering the planet's interior heating or instellation could provide more physically-motivated constraints on delayed cooling.
In the particular case of late arrivals, though the evolution of a hot Jupiter's orbit in the era of its parent protoplanetary disk can impact its subsequent thermal and radius evolution \citep{MolLous20}, high-eccentricity tidal migration can effectively ``reset'' the thermal evolution of a late-arriving hot Jupiter that cooled and contracted far from its host star for over a Gyr \citep[e.g.,][]{Liu08, Ibgui09}.
On the other end of the exoplanet age distribution, a larger sample of hot Jupiters orbiting evolved stars could also help constrain the timescale for reinflation and therefore the effects of deep heating thanks to their much faster rate of stellar evolution.
These larger and more precise data sets would enable further characterization of the mechanisms that heat hot Jupiters, leading to much more informed constraints on individual mechanism contributions as well as their typical impacts on observed planetary atmosphere characteristics.

\section{Conclusions} \label{sec:Conclusions}
We have shown evidence for dominant shallow heating in hot Jupiter interiors based on a drastic reduction in the rate of outward energy flow.
To accomplish this, we updated our giant planet interior structure and thermal evolution model to reduce the intrinsic luminosity from a planet's deep interior while simultaneously lowering the deep heating power to keep the planet's interior in thermal equilibrium (see Section \ref{sec:thermalEvolutionModel}).
We applied these models to precise and accurate archival photometry and astrometry-derived stellar and planetary system properties (see Section \ref{sec:stellarInference}) in a hierarchical Bayesian framework (see Section \ref{sec:hierarchicalBayesianInference}), finding that the population's intrinsic luminosities were reduced by 95-97\% relative to models with exclusively deep heating \citep[e.g.][]{Guillot2002, Thorngren2018}.
As our model uses an updated equation of state in comparison to \citet{Thorngren2018}, we allowed for an offset between values of the typical heating experienced by hot Jupiters.
While we did find a small nonzero value for this offset, it is uncorrelated with the intrinsic luminosity reduction factor across a range of modeled initial specific entropies, indicating that our findings were not biased by it.

Our findings are best explained by the presence of shallow heating mechanisms that take place near or just below the radiative--convective boundary.
This reconciles a tension between deep heating models that explain radius reinflation with delayed-cooling models that explain young highly-inflated hot Jupiters.
Because shallow heating mechanisms like advection or Ohmic dissipation are coupled with atmospheric winds, we expect that phase curve and other atmospheric circulation-related observables should inherit trends in heating efficiency.
Because shallow heating mechanisms play a major role in both the thermal structure and atmospheric flows of hot Jupiters, they will be important to consider in global circulation models (GCMs) used to study these objects.

\begin{acknowledgments}
We thank the anonymous referee for their helpful comments.
Stephen P. Schmidt is supported by the National Science Foundation Graduate Research Fellowship Program under Grant No. DGE2139757.
This material is based on work supported by the National Aeronautics and Space Administration under grant No. 80NSSC23K0266 issued through the Exoplanets Research Program (XRP).
This work has made use of data from the European Space Agency
(ESA) mission Gaia (\url{https://www.cosmos.esa.int/gaia}),
processed by the Gaia Data Processing and Analysis Consortium (DPAC,
\url{https://www.cosmos.esa.int/web/gaia/dpac/consortium}).  Funding for
the DPAC has been provided by national institutions, in particular
the institutions participating in the Gaia Multilateral Agreement.
This publication makes
use of data products from the Two Micron All Sky Survey, which is
a joint project of the University of Massachusetts and the Infrared
Processing and Analysis Center/California Institute of Technology,
funded by the National Aeronautics and Space Administration and the
National Science Foundation.  This publication makes use of data products
from the Wide-field Infrared Survey Explorer, which is a joint project
of the University of California, Los Angeles, and the Jet Propulsion
Laboratory/California Institute of Technology, funded by the National
Aeronautics and Space Administration.  This research has made use of
the SIMBAD database, operated at CDS, Strasbourg, France \citep{wen00}.
This research has made use of the VizieR catalog access tool, CDS,
Strasbourg, France.  The original description of the VizieR service
was published in \citet{och00}.  This research has made use of NASA's
Astrophysics Data System Bibliographic Services.

This publication makes use of data products from SkyMapper.
The national facility capability for SkyMapper has been funded through ARC
LIEF grant LE130100104 from the Australian Research Council, awarded to
the University of Sydney, the Australian National University, Swinburne
University of Technology, the University of Queensland, the University
of Western Australia, the University of Melbourne, Curtin University of
Technology, Monash University and the Australian Astronomical Observatory.
SkyMapper is owned and operated by The Australian National University's
Research School of Astronomy and Astrophysics.  The survey data were
processed and provided by the SkyMapper Team at ANU.  The SkyMapper node
of the All-Sky Virtual Observatory (ASVO) is hosted at the National
Computational Infrastructure (NCI).  Development and support of the
SkyMapper node of the ASVO has been funded in part by Astronomy Australia
Limited (AAL) and the Australian Government through the Commonwealth's
Education Investment Fund (EIF) and National Collaborative Research
Infrastructure Strategy (NCRIS), particularly the National eResearch
Collaboration Tools and Resources (NeCTAR) and the Australian National
Data Service Projects (ANDS).

This publication makes use of data products from the Sloan Digital Sky Survey II (SDSS-II).
Funding for the SDSS and SDSS-II has
been provided by the Alfred P. Sloan Foundation, the Participating
Institutions, the National Science Foundation, the U.S. Department of
Energy, the National Aeronautics and Space Administration, the Japanese
Monbukagakusho, the Max Planck Society, and the Higher Education Funding
Council for England. The SDSS Web Site is \url{http://www.sdss.org/}.
The SDSS is managed by the Astrophysical Research Consortium for the
Participating Institutions. The Participating Institutions are the
American Museum of Natural History, Astrophysical Institute Potsdam,
University of Basel, University of Cambridge, Case Western Reserve
University, University of Chicago, Drexel University, Fermilab, the
Institute for Advanced Study, the Japan Participation Group, Johns
Hopkins University, the Joint Institute for Nuclear Astrophysics, the
Kavli Institute for Particle Astrophysics and Cosmology, the Korean
Scientist Group, the Chinese Academy of Sciences (LAMOST), Los Alamos
National Laboratory, the Max-Planck-Institute for Astronomy (MPIA),
the Max-Planck-Institute for Astrophysics (MPA), New Mexico State
University, Ohio State University, University of Pittsburgh, University
of Portsmouth, Princeton University, the United States Naval Observatory,
and the University of Washington.

This publication makes use of data products from the Sloan Digital Sky Survey III (SDSS-III).
Funding for SDSS-III has been provided
by the Alfred P. Sloan Foundation, the Participating Institutions, the
National Science Foundation, and the U.S. Department of Energy Office of
Science. The SDSS-III web site is http://www.sdss3.org/.  SDSS-III is
managed by the Astrophysical Research Consortium for the Participating
Institutions of the SDSS-III Collaboration including the University
of Arizona, the Brazilian Participation Group, Brookhaven National
Laboratory, Carnegie Mellon University, University of Florida, the French
Participation Group, the German Participation Group, Harvard University,
the Instituto de Astrofisica de Canarias, the Michigan State/Notre
Dame/JINA Participation Group, Johns Hopkins University, Lawrence
Berkeley National Laboratory, Max Planck Institute for Astrophysics,
Max Planck Institute for Extraterrestrial Physics, New Mexico State
University, New York University, Ohio State University, Pennsylvania
State University, University of Portsmouth, Princeton University, the
Spanish Participation Group, University of Tokyo, University of Utah,
Vanderbilt University, University of Virginia, University of Washington,
and Yale University.

This publication makes use of data products from the Sloan Digital Sky Survey IV (SDSS-IV).
Funding for the Sloan Digital Sky Survey IV has
been provided by the Alfred P. Sloan Foundation, the U.S.  Department
of Energy Office of Science, and the Participating Institutions.
SDSS-IV acknowledges support and resources from the Center for High
Performance Computing  at the University of Utah. The SDSS website is
\url{www.sdss4.org}.  SDSS-IV is managed by the Astrophysical Research
Consortium for the Participating Institutions of the SDSS Collaboration
including the Brazilian Participation Group, the Carnegie Institution
for Science, Carnegie Mellon University, Center for Astrophysics |
Harvard \& Smithsonian, the Chilean Participation Group, the French
Participation Group, Instituto de Astrof\'isica de Canarias, The Johns
Hopkins University, Kavli Institute for the Physics and Mathematics of
the Universe (IPMU) / University of Tokyo, the Korean Participation Group,
Lawrence Berkeley National Laboratory, Leibniz Institut f\"ur Astrophysik
Potsdam (AIP),  Max-Planck-Institut f\"ur Astronomie (MPIA Heidelberg),
Max-Planck-Institut f\"ur Astrophysik (MPA Garching), Max-Planck-Institut
f\"ur Extraterrestrische Physik (MPE), National Astronomical Observatories
of China, New Mexico State University, New York University, University of
Notre Dame, Observat\'ario Nacional / MCTI, The Ohio State University,
Pennsylvania State University, Shanghai Astronomical Observatory,
United Kingdom Participation Group, Universidad Nacional Aut\'onoma
de M\'exico, University of Arizona, University of Colorado Boulder,
University of Oxford, University of Portsmouth, University of Utah,
University of Virginia, University of Washington, University of Wisconsin,
Vanderbilt University, and Yale University.

This publication makes use of data products from the Pan-STARRS1 Survey.
The Pan-STARRS1 Surveys
(PS1) and the PS1 public science archive have been made possible
through contributions by the Institute for Astronomy, the University
of Hawaii, the Pan-STARRS Project Office, the Max-Planck Society and
its participating institutes, the Max Planck Institute for Astronomy,
Heidelberg and the Max Planck Institute for Extraterrestrial Physics,
Garching, The Johns Hopkins University, Durham University, the University
of Edinburgh, the Queen's University Belfast, the Harvard-Smithsonian
Center for Astrophysics, the Las Cumbres Observatory Global Telescope
Network Incorporated, the National Central University of Taiwan, the
Space Telescope Science Institute, the National Aeronautics and Space
Administration under Grant No. NNX08AR22G issued through the Planetary
Science Division of the NASA Science Mission Directorate, the National
Science Foundation Grant No. AST-1238877, the University of Maryland,
Eotvos Lorand University (ELTE), the Los Alamos National Laboratory,
and the Gordon and Betty Moore Foundation.

\end{acknowledgments}

\vspace{5mm}
\facilities{CDS, CTIO:2MASS, Exoplanet Archive, FLWO:2MASS, Gaia, GALEX, IRSA, NEOWISE,
PS1, Skymapper, Sloan, WISE}

\software{\texttt{astropy} \citep{AstropyI, AstropyII, AstropyIII},
\texttt{dustmaps} \citep{gre18},
\texttt{gaiadr3\_zeropoint} \citep{lin21a},
\texttt{isochrones} \citep{mor15},
\texttt{matplotlib} \citep{hunter2007matplotlib},
\texttt{numpy} \citep{harris2020array},
\texttt{pandas} \citep{McKinney_2010, reback2020pandas},
\texttt{R} \citep{r25},
\texttt{scipy} \citep{jones2001scipy,2020SciPy-NMeth},
\texttt{TOPCAT} \citep{tay05}}

\bibliographystyle{aasjournalv7}
\bibliography{article_bibliography}{}
\end{document}